\documentclass[twocolumn,runningheads]{svjour2}

\smartqed  % flush right qed marks, e.g. at end of proof

\usepackage{graphicx}

\journalname{Astrophysics and Space Science}

\begin{document}

\title{Population Studies of the Unidentified EGRET Sources\thanks{This work was supported by the Kavli Institute for Cosmological Physics through the grant NSF PHY-0114422 and by DOE grant DE-FG0291-ER40606 at the University of Chicago.}
}

\author{Jennifer M. Siegal-Gaskins\and
	  Vasiliki Pavlidou\and
        Angela V. Olinto\and
	  Carolyn Brown\and
	Brian D. Fields
}

\institute{J. M. Siegal-Gaskins \and V. Pavlidou \and A. V. Olinto \and C. Brown\at
University of Chicago\\
Chicago, IL 60637 \\
\email{gaskins@oddjob.uchicago.edu}\\
\and
B. D. Fields\at
University of Illinois at Urbana-Champaign\\
Urbana, IL 61801 \\
}

\date{Received: date / Accepted: date}
% The correct dates will be entered by the editor

\maketitle

\begin{abstract}

The third EGRET catalog contains a large number of unidentified sources.  This
subset of objects is expected to include known gamma-ray emitters of Galactic origin 
such as pulsars and supernova remnants, in addition to an extragalactic population
of blazars.
However, current data allows the intriguing possibility that some of these objects
may represent a new class of yet undiscovered gamma-ray sources.  Many 
theoretically motivated candidate emitters (e.g.\ clumps of annihilating dark matter 
particles) have been suggested to account for these detections.
We take a new approach to determine to what extent this population is Galactic
and to investigate the nature of the possible Galactic component.  
By assuming that galaxies similar to the Milky Way should host 
comparable populations of objects, 
we constrain the allowed Galactic abundance and distribution of various 
classes of gamma-ray sources
using the EGRET data set. 
We find it is highly improbable that a large number of the unidentified sources are members of a Galactic halo population, but that a distribution of the sources entirely in the disk and bulge is plausible.
Finally, we discuss the additional constraints and new insights that GLAST will 
provide.

\end{abstract}

\section{Introduction}
\label{intro}

The Energetic Gamma Ray Experiment Telescope (EGRET) measured the gamma-ray emission at energies greater than 100 MeV across the entire sky.  At the time the third EGRET catalog~\cite{hartman_etal_99} was published, unidentified sources accounted for more than half of the reported detections.  Since then only a handful of those sources have been associated with known low-energy counterparts.  
Theoretical work has produced many candidate sources for these detections, including known Galactic source classes not previously confirmed as gamma-ray emitters (e.g.\ microquasars~\cite{paredes_etal_00}) and newly proposed populations of high energy galactic sources (e.g.\ annihilating dark matter clumps~\cite{berezinsky_etal_03,bergstrom_etal_99,blasi_olinto_tyler_03,calcaneoroldan_moore_00,tasitsiomi_olinto_02,taylor_silk_03,ullio_etal_02} and intermediate mass black holes~\cite{bertone_etal_05}).  Meanwhile, multiwavelength searches continue to look for counterparts in confirmed gamma-ray emitting classes (e.g.\ blazars, pulsars, and supernova remnants).

However, the sources in the third EGRET catalog typically have larger error boxes than surveys in lower frequencies, making identification with known Galactic sources by positional coincidence difficult~\cite{gehrels_michelson_99,lapalombara_etal_06}.  The EGRET data set provides spectral and variability information which can be used to strengthen an identification (and has proved useful, e.g.~in identifying pulsar wind nebulae~\cite{roberts_etal_05}), but in some cases this information is not available or subject to large uncertainties (see variability analysis in~\cite{nolan_etal_03}).  Future experiments which can probe smaller variability timescales may be able to use this information to identify binaries and pulsars.

The situation is less favorable for identifying sources as members of theoretically-motivated classes of gamma-ray emitters without previously established detections.  The uncertainties inherent in the emission features of unconfirmed source classes make identification of individual sources tentative at best.
While emission from dark matter annihilation is predicted to have a unique signature (a `smoking gun' spectral line), for many scenarios it is unlikely that even next-generation experiments could detect such a signal.  

In light of recent theoretical developments, and with the scheduled launch of the Gamma-ray Large Area Space Telescope (GLAST) mission next year, this is an ideal time to revisit the EGRET data set.  The number of detected gamma-ray sources will be increased substantially by GLAST, and with it the number of sources to be identified with known low-energy counterparts.  With GLAST's expanded source catalog, following each detection with multiwavelength observations will almost certainly be an impractical method for making identifications.  While other gamma-ray telescopes currently in operation such as HESS, MAGIC, and VERITAS will be able to use their superior angular resolution and variability measurements to help identify sources, they face similar issues with regard to the burden of multiwavelength follow-up observations of large numbers of objects not easily identified by other methods.

This study statistically investigates the nature of the EGRET unidentified sources.  
By assuming that the nearby galaxy M31 should host gamma-ray emitting populations similar to those of the Milky Way, we determine whether the unidentified sources can be of Galactic origin and place constraints on the spatial distribution of this population without assuming a specific candidate emitter.  Our approach uses only angular position and flux information to evaluate the plausibility of Galactic distributions of these sources. 
 
In \S\ref{constraints} we discuss the unidentified sources, outline our approach for placing constraints on unidentified source populations, and present our results.  We comment on the new insights GLAST will bring in \S\ref{glast} and conclude in \S\ref{conc}.

\section{Constraining a Galactic Population with M31}
\label{constraints}

Like the Milky Way, M31 is a luminous, high surface brightness galaxy with a mass of $\sim10^{12}$ M$_{\odot}$.  Having formed and evolved in a similar environment,
M31 is also akin to the Milky Way in its structure and dynamical properties (see e.g.~\cite{kzs_02}).
Because of this close resemblance, it seems likely that any gamma-ray emitting source populations found in the Milky Way are also present in M31.  Under this assumption we test candidate Galactic populations by asking whether a comparable population in M31 is consistent with current observational constraints.

EGRET did not detect M31, determining only an upper limit to its flux of F$_{\rm M31}(>$100 MeV$)<1.6\times 10^{-8}$ cm$^{-2}$ s$^{-1}$~\cite{blom_etal_99}.  We take the distance to M31 to be 670 kpc~\cite{blom_etal_99}, which gives an upper limit on the luminosity of M31 of $8.6\times10^{41}$ s$^{-1}$ (or 0.090 (kpc/cm)$^{2}$ s$^{-1}$) for energies greater than 100 MeV.

However, the entire gamma-ray luminosity of M31 cannot be attributed to point sources.  Pavlidou and Fields~\cite{pavlidou_fields_01} calculate the expected flux of diffuse gamma-ray emission due to cosmic ray interactions to be F$_{\rm diff}(>$100 MeV$)\sim1.0\times10^{-8}$ cm$^{-2}$ s$^{-1}$, almost two thirds of the EGRET upper limit.  Consequently, the expected luminosity of M31 due to point sources is restricted to less than $\sim3.2\times10^{41}$ s$^{-1}$ (or 0.034 (kpc/cm)$^{2}$ s$^{-1}$).  We use these luminosity limits to constrain the Galactic distribution of the unidentified sources by requiring that the luminosity of the distribution be consistent with these bounds.  We note that the upper limit on the total gamma-ray luminosity is robust; the luminosity upper limit which excludes the diffuse component is, however, model-dependent, but we include it for completeness.
 
Emission from unresolved point sources is another component of the Galactic luminosity which is measured as diffuse emission.  The contribution from unresolved sources may be relevant because our assumption that M31 and the Milky Way host similar populations of gamma-ray emitting sources requires that unresolved sources enhancing the Galactic diffuse emission also be present in M31.  The prediction we use for the diffuse flux of M31 only represents the expected genuinely diffuse emission, and does not account for the luminosity enhancement by unresolved point sources.  However, due to uncertainties in the magnitude of this emission, we again make the conservative choice to ignore this guaranteed component when placing constraints, allowing all of the expected point source luminosity to be attributed to the resolved unidentified sources.

Additionally, there is already a guaranteed contribution to the luminosity of point sources from confirmed detections of members of known Galactic source classes, which further reduces the luminosity available to the unidentified sources.  From the EGRET data and other observational efforts, a number of Galactic gamma-ray emitting objects such as pulsars and supernova remnants have been identified.  
However, the total gamma-ray luminosity of these objects is very small 
compared to the diffuse luminosity of either the Milky Way or M31.
Including this component would not alter our results in any substantial way, so for
simplicity we ignore it.

Throughout this work we consider the unidentified sources to be members of a class (or classes) of Galactic gamma-ray emitters with a predicted spatial distribution.  We do, however, expect that our catalog contains some sources that do not meet this criterion (in particular extragalactic sources which do not contribute to the point source luminosity of the Milky Way).  While contamination by extragalactic objects would lead us to overestimate the Galactic luminosity of a test population by increasing the number of sources treated as Galactic, constraints on the abundance of Galactic sources with the assumed distribution would not be invalidated.  
Ultimately, we seek to calculate the maximal fraction of unidentified
sources which could be members of a particular Galactic population. As a
first step, we test in this work whether
all unidentified sources could be members of a population (or populations) with an assumed spatial distribution.
Constraining the abundance of Galactic sources following a given spatial distribution allows us to draw conclusions about the global properties of any proposed Galactic gamma-ray emitting population, rather than testing the plausibility of a single population.

\subsection{The Unidentified Sources}
\label{sources}

The third EGRET catalog contains 271 sources (E$>$100 MeV), 101 of which were
initially identified by the EGRET team.  The majority of those sources were identified as blazars, along with a smaller number of pulsars, the Large Magellanic
Cloud, a radio galaxy, and a solar flare.  An additional 43 of the originally unidentified sources subsequently have been suggested to be associated with plausible counterparts,
including binaries, supernova remnants, gas clouds, microquasars, and black holes.

We use an updated listing of EGRET unidentified
sources\footnote{Listing compiled and actively maintained by C.~Brown, and available at\\ http://home.uchicago.edu/carolynb/unidentified\_sources}, excluding all sources for which an identification has been suggested, regardless of the significance of the identification.  It is important to note that the identifications included in this compilation are simply the results reported in recent publications; the validity of these suggested counterparts has not been evaluated by a single standard.  Similar to ignoring the expected luminosity contributions described in the previous section, omitting these sources with suggested counterparts from our study is a conservative choice because increasing the number of sources in our candidate population would lead to an increase in the total population luminosity, and hence the realization would be more likely to violate the M31 luminosity constraints.  Of course, these excluded sources are gamma-ray emitters and in some cases have been suggested to be of Galactic origin, so they would fall into the category of confirmed Galactic detections as discussed above.  
Since these associations
are in many cases tentative, the distances determined for the sources
are uncertain and can only be treated as estimates. 
From the estimated distances of the sources with recently suggested counterparts, 
the luminosity of these sources is $\sim 10^{39}$ s$^{-1}$, more than 2 orders of
magnitude lower than the EGRET upper limit for the gamma-ray luminosity
of M31.
Consequently, we ignore this contribution to the Galactic luminosity, allowing the candidate source population under consideration to contribute the entire point source luminosity.

It is interesting to note that many of the unidentified sources located in the Galactic plane are among those recently suggested to be associated with a low energy counterpart.  As a result, the sky distribution of unidentified sources appears more isotropic, and the overdensity of sources near the Galactic center is less pronounced.  However,  
testing whether the subset of sources  
we use in this study is consistent with the presence of a dominant
isotropic component requires a detailed statistical treatment taking
into account variations in the EGRET exposure map, and is beyond the scope of this work.

For this study we use the fluxes measured by EGRET, along with the angular positions of the sources.
Each source in the catalog is also reported with an estimate of positional uncertainty, $\Theta_{95}$, typically $\sim$1$^\circ$.  
This value is for most objects the angular radius of a circle containing the same solid angle as the 95\% confidence level contour.  
However, in general the contour is not a circle but rather a complex shape, and so for simplicity we will consider $\Theta_{95}$ to be the angular
uncertainty of both the Galactic longitude $\ell$ and latitude $b$ measurements.  We note that Mattox {\it et al.}~\cite{mattox_etal_01} have provided elliptical contour fits to the EGRET sources, but  
because we do not expect the mass density to vary significantly over such small scales, 
generalizing the error contours to squares is a suitable approximation for our purposes.

\subsection{Assigning Source Distances}
\label{distances}

The EGRET data provides a measurement of the flux and the angular position of each source, so the distance to each unidentified source is needed to calculate the total luminosity of a candidate population.  Our method for assigning distances is motivated by the goal of testing proposed source populations, so we determine the distance for each source by considering the expected Galactic distribution of the population.  

In general, candidate Galactic populations can be classified as living either in the halo (e.g.\ dark matter clumps, intermediate mass black holes) or in the disk and bulge (e.g.\ pulsars, supernova remnants, and other baryonic objects).  Because these populations can be associated with a measured mass component of the Galaxy, we use mass density as a proxy for source density.  Using a Monte Carlo algorithm we then generate realizations of Galactic source distributions and calculate the total luminosity of the population for each run.

For each angular source position ($\ell,b$) we construct a cumulative distribution function for a given Galactic mass distribution to describe the likelihood of the source being within a particular distance along that line of sight.  The probability of a source with a given angular position being located within a distance $d$ from us is then given by

\begin{equation}
\mathcal{P}(d)=\frac{M_{\rm encl}(d)}{M_{encl}(d_{\rm max})},
\end{equation}

where $d_{\rm max}$ is the distance along that line of sight at which we truncate our mass distribution to produce a finite volume in which to place the sources.

The mass enclosed along the line of sight within a solid angle    
defined by the positional error boxes and extending from the observer
out to a distance $d$ is given by

\begin{equation}
M_{\rm encl}(d) = \int_{0}^{d} \int_{\ell_{-}}^{\ell_{+}}  \int_{b_{-}}^{b_{+}} \rho(z,\ell,b) z^{2} \cos(b) db d\ell dz,
\end{equation}

where $\ell_{\pm}=\ell \pm \Theta_{95}$, $b_{\pm}=b \pm \Theta_{95}$, and $\rho$ is the mass density.  The variable $z$ describes integration along the line of sight.

\subsection{Results}
\label{results}

We first consider candidate populations correlated with the dark matter distribution of the galaxy.  We model the dark matter halo of the Milky Way using the density profile proposed by Navarro, Frenk, and White (NFW profile)~\cite{nfw_95} with mass M$_{\rm halo} = 10^{12}$ M$_{\odot}$ and concentration $c = 12$~\cite{kzs_02}, truncating the profile at a radius of 100 kpc from the Galactic Center.  For this scenario, we include only unidentified sources with $|b|>5^\circ$, omitting the 26 sources which appear to be in the Galactic plane.

For the case of source populations expected to reside in the disk and bulge, we approximate the mass distribution with a Miyamoto-Nagai disk~\cite{miyamoto_nagai_75} and a Hernquist bulge~\cite{hernquist_90}.  We take the disk mass to be M$_{\rm d} = 4 \times 10^{10} $ M$_{\odot}$~\cite{kzs_02}, and the scale parameters $a$ and $b$ to be 6.5 kpc and 0.26 kpc respectively~\cite{johnston_etal_96}.  For the bulge mass we use M$_{\rm b} = 8.0 \times 10^{9} $ M$_{\odot}$ ($m_{1} + m_{2}$ in~\cite{kzs_02}), with the scale parameter $a=0.7$ kpc~\cite{johnston_etal_96}.  The disk and bulge profiles are truncated at a radius of 30 kpc.  We test this scenario first using only the sources which appear to be in the Galactic plane ($|b|<5^\circ$), and then considering all sources. 

We use a Monte Carlo algorithm to generate realizations of these population distributions.  For each realization
a distance is assigned to each source by sampling the appropriate cumulative distribution function associated with the angular position of that source.  The total luminosity for the realization is then calculated.

\begin{figure}
	\centering
	\includegraphics[width=.95\columnwidth]{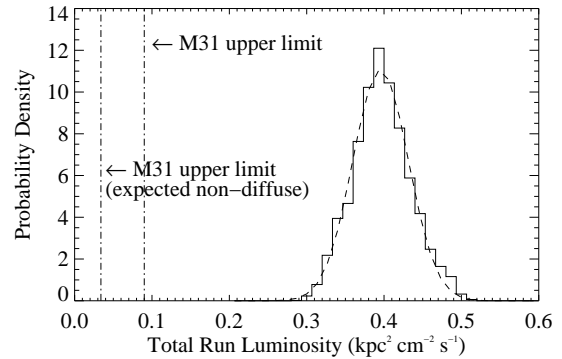}
	\caption{Probability density function for total luminosity for a halo population of sources.  The histogram shows the total luminosities for 5000 runs, using sources with $|b|>5^\circ$.   The M31 total luminosity upper limit and derived upper limit for the luminosity of point sources are shown for reference.  See text for details.}
	\label{fig:haloresults}
\end{figure}

Figure \ref{fig:haloresults} shows the distribution of the total source luminosity for a halo population using 5000 realizations.  Assuming a normal Gaussian distribution, the central value and width are $\mu = 0.396$ and $\sigma = 0.0360$ respectively, in units of (kpc/cm)$^{2}$ s$^{-1}$.  The central value of the total luminosity distribution is more than a factor of 4 greater than the observational luminosity limit for M31, and more than an order of magnitude greater than the expected upper limit for point sources.  The possibility of any halo distribution of these unidentified sources producing a luminosity below even the maximum observational limit is ruled out at extremely high confidence, indicating that EGRET did not detect a significant number of Galactic halo objects, such as annihilating dark matter clumps or intermediate mass black holes.  However, for a halo distribution our study considers only the case that the entire set of unidentified sources with $|b|>5^\circ$ are members of the halo population, so additional tests would be needed to determine how large a subset of these sources could exist in the halo without exceeding the M31 constraint.

\begin{figure}
	\centering
	\includegraphics[width=.95\columnwidth]{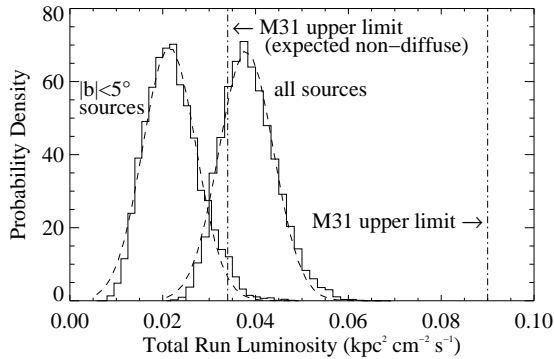}
	\caption{Probability density function for total luminosity for a population of sources associated with the disk and bulge.  The resulting distribution for 5000 runs using only sources with $|b|<5^\circ$ is shown along with the corresponding distribution when all sources are used.  As in Figure \ref{fig:haloresults}, the M31 limits are marked for reference.}
	\label{fig:dbresults}
\end{figure}

Figure \ref{fig:dbresults} shows the results for disk and bulge distributions.  We produce 5000 realizations for each set of sources, first using only the sources which appear to be located in the Galactic plane ($|b|<5^\circ$) and then using all sources.  Fitting these distributions to a normal Gaussian gives $\mu=0.0214$ and $\sigma=0.00575$ for the Galactic plane sources only, and  $\mu=0.0380$ and $\sigma=0.00583$ for all sources, again in units of (kpc/cm)$^{2}$ s$^{-1}$.  For both cases the observational luminosity upper limit lies comfortably above the distributions.  For the Galactic plane sources only, we find a 99\% likelihood that the total luminosity for a realization will be smaller than the expected point source upper limit, and a 25\% likelihood if all of the sources are considered.  In light of the uncertainties in our assumption that the population of gamma-ray emitting point sources in M31 is comparable to that of the Milky Way, our luminosity test is consistent with the possibility that all of the unidentified sources are part of a Galactic population distributed in the disk and bulge.  

Qualitatively, we would expect that a halo population would yield a higher
total luminosity than a disk and bulge population, and would therefore be
more likely to violate the M31 constraint. Members of a halo population
tend to be located further away from us than members of a
disk and bulge population, consequently an unidentified source of   
a given gamma-ray flux is typically assigned a higher individual
luminosity if it is assumed to reside in the halo as opposed to the disk and bulge. As a result, the summed luminosity of unidentified sources for the case of a halo population is consistently higher than the summed luminosity of the same sources for the case of a disk and bulge population.

\section{New Insights from GLAST}
\label{glast}

The GLAST mission is scheduled to launch in late 2007.  With a point source sensitivity of less than $6 \times 10^{-9}$ cm$^{-2}$ s$^{-1}$ (more than an order of magnitude more sensitive than EGRET)\footnote{See http://glast.gsfc.nasa.gov and references therein.}, GLAST will provide a wealth of new information about the gamma-ray sky.

Of particular interest to this study is the likely detection of M31~\cite{digel_etal_00,pavlidou_fields_01}.  A measurement of the M31 gamma-ray flux will significantly narrow the range of plausible point source luminosities, and provide stronger constraints on the properties of new Galactic gamma-ray emitting populations.  Additionally, GLAST may be able to provide evidence of similarities in the gamma-ray emission of the Milky Way and M31, which would strengthen the assumption underlying this work.
A determination of the M31 flux would also test the validity of the predicted diffuse gamma-ray emission.

GLAST will also provide a new determination of the Galactic and extragalactic gamma-ray backgrounds.  Measured background emission includes both genuinely diffuse emission and emission from unresolved sources.  Because of GLAST's higher sensitivity, it will resolve many more sources than EGRET, and consequently those sources will no longer contribute to the background emission.  As discussed in Pavlidou {\it et al.}~\cite{pavlidou_etal_06}, the decreases in the Galactic and extragalactic components of the background emission measured by GLAST will indicate to what extent the newly resolved sources are of Galactic or extragalactic origin.  In this way, GLAST will constrain the abundance of faint Galactic source populations.

The expanded source catalog produced by GLAST will greatly improve the robustness of isotropy studies.  Isotropy could be a powerful tool for evaluating the likelihood of a population being of Galactic origin, but it is difficult to make statistically significant statements using the EGRET catalog alone due to the limited number of sources.  The issue is further complicated by EGRET's variation in flux sensitivity in different areas of the sky, which strongly favors sources in the Galactic plane.  By greatly increasing the number of detections, GLAST will be better able to disentangle the Galactic and extragalactic source populations.

In addition to lower flux sensitivity, GLAST is expected to achieve much smaller angular resolution than EGRET, which will greatly aid in identifications by positional coincidence and reduce source confusion.  Recently, Casandjian {\it et al.}~\cite{casandjian_05} reanalyzed the EGRET data using a new interstellar emission model and found that several new sources appeared, while many unidentified sources were no longer detected at high significance.  This study suggests that source confusion may have been an important limitation of the EGRET data set.

By incorporating spectral and variability information, the larger number of sources detected by GLAST and identified may be used to better determine defining characteristics of different source classes.  Using this information to suggest likely source types for unidentified sources would be useful for designing candidate counterpart searches.  

Finally, applying the methods of this study and other statistical techniques to probe the nature of the detected sources may prove useful when dealing with the large data set GLAST is expected to generate.

\section{Conclusions}
\label{conc}

We used the observational upper limit for the gamma-ray luminosity of M31 along with the assumption that M31 and the Milky Way host similar gamma-ray emitting populations to constrain the allowed spatial distribution of the unidentified sources in the third EGRET catalog.  

We find that it is highly unlikely that a substantial fraction of the unidentified sources are members of a Galactic halo population.  This result implies that EGRET's unidentified sources do not consist primarily of annihilating dark matter clumps or intermediate mass black holes. 

For the case of source populations expected to be associated with the disk and bulge, we find that all of the sources can be of Galactic origin without exceeding our assumed luminosity upper limit.  
However, satisfying the luminosity constraint alone does not guarantee that a proposed distribution is plausible.  Any realistic Galactic population is expected to be distributed according to symmetries of the Galaxy (e.g.\ it is unlikely that a population would cluster near our position).  Our method of assigning distances effectively projects all of the sources into the disk and bulge or into the halo.  Particularly in the case of a population associated with the disk and bulge, careful isotropy studies taking into account EGRET's sensitivity map would be needed to determine whether realizations of this population represent plausible spatial distributions.

It is important to keep in mind that although we expect M31 and the Milky Way to be similar in gamma-rays, clearly we do not expect them to be identical.  Depending on the characteristics of the source population and the properties of M31, we expect that the total population luminosities we calculated by placing the unidentified sources in the Milky Way could vary by a factor of a few for a corresponding population in M31.  Although a variation of this size would not change our conclusion that a halo population is highly unlikely, it could easily shift the luminosity distribution for a population in the disk and bulge above the observational M31 limit.  

In addition, there is currently only an upper limit to the M31 flux, so a flux measurement would necessarily affirm or tighten the luminosity constraint.  Once a determination of the M31 flux is made, it is possible that no new Galactic populations could be accommodated.  Furthermore, a flux determination would also suggest a lower limit to the point source flux.  With current data this study cannot require that any of the sources be Galactic, leaving open the possibility that all of the unidentified sources are blazars or other extragalactic objects.

We did not take into account a number of probable contributors to the Galactic luminosity: diffuse emission from unresolved point sources, identified Milky Way objects, and sources with recently suggested low-energy counterparts.  We expect that M31 will have similar populations of these Galactic sources, so in principle this will decrease the allowed luminosity of new proposed populations, and consequently further constrain their abundance and distribution.

The new information GLAST brings, with respect to both known gamma-ray sources and possible detections of proposed emitters, will help to determine the nature of the unidentified sources.  Statistical techniques such as those employed in this study will be useful in making meaningful statements about gamma-ray emitting populations when large numbers of individual identifications are not feasible.

\begin{acknowledgements}
\emph{We are grateful to A.\ Strong and O.\ Reimer for their insightful comments and discussions relating to this work.}
\end{acknowledgements}

% BibTeX users please use
\bibliographystyle{spmpsci}
\bibliography{bibliography_bcnedit}   % name your BibTeX data base

\end{document}